\documentclass[twoside]{article}

\usepackage[accepted]{aistats2019mod}
\usepackage[round]{natbib}

\usepackage{graphicx}
\usepackage{footnote}
\usepackage[colorlinks,linkcolor=black,citecolor=blue,urlcolor=blue,filecolor=blue]{hyperref}
\usepackage{microtype}
\usepackage{amssymb}

\usepackage{blkarray}

\begin{document}

\twocolumn[

\aistatstitle{Variable selection for Gaussian processes via sensitivity analysis of the posterior predictive distribution}

\aistatsauthor{ Topi Paananen \And Juho Piironen \And  Michael Riis Andersen \And Aki Vehtari }

\aistatsaddress{ {\tt topi.paananen@aalto.fi} \And {\tt juho.piironen@aalto.fi} \And {\tt michael.riis@gmail.com} \And {\tt aki.vehtari@aalto.fi} } 
\vspace{-1.6em}
\aistatsaddress{ Helsinki Institute for Information Technology, HIIT \\ Aalto University, Department of Computer Science }

]

\begin{abstract}
  Variable selection for Gaussian process models is often done using automatic relevance determination, which uses the inverse length-scale parameter of each input variable as a proxy for variable relevance.
  This implicitly determined relevance has several drawbacks
  that prevent the selection of optimal input variables in terms of predictive performance.
  To improve on this, we propose two novel variable selection methods for Gaussian process
  models that utilize the predictions of a full model in the vicinity of
  the training points and thereby rank the variables based on their
  predictive relevance.
  Our empirical results on synthetic and real world data sets
  demonstrate improved variable selection compared to
  automatic relevance determination in terms of variability and predictive performance.
\end{abstract}

\section{INTRODUCTION}

Often
the goal of supervised learning is not only to learn
the relationship between the predictors and target variables, but
to also assess the predictive relevance of the input variables.
A relevant input variable is one with a high predictive power
on the target variable~\citep{vehtari2012survey}.
In many applications, simplifying a model by selecting only the most relevant input variables is important for two reasons.
Firstly, it makes the model more interpretable and understandable by domain experts.
Secondly, it may reduce future costs
if there is a price associated with measuring or predicting with many variables.
Here, we focus on methods that select a subset of the original variables, as opposed to constructing
new features, as this preserves the interpretability of the variables.

Gaussian processes (GPs) are flexible, nonparametric models
for regression and classification in the Bayesian framework~\citep{rasmussen2006gaussian}.
The relevance of input variables of a fitted GP model is often inferred
implicitly from the length-scale
parameters of the GP covariance function.
This is called automatic relevance determination (ARD), a term that originated in the neural network literature~\citep{mackay1994bayesian,neal1995bayesian}, and since then has been used
extensively for both Gaussian processes~\citep{williams1996gaussian,seeger2000bayesian} and other models~\citep{tipping2000relevance,wipf2008new}.

Alternative to ARD, variable selection via sparsifying spike-and-slab priors is possible also with
Gaussian processes~\citep{linkletter2006variable,savitsky2011variable}.
The drawback of these methods is that they
require using a Markov chain Monte Carlo method for inference, which is computationally expensive with Gaussian processes.
Due to space constraints, we will not consider sparsifying priors in this study.
The predictive projection method originally devised for generalized linear models~\citep{goutis1998model,dupuis2003variable}
has also been implemented for Gaussian processes~\citep{piironen2016projection}.
This method can potentially select variables with good predictive performance, but
has a substantial computational cost due to the required exploration of the model space.

Due to the close connection between Gaussian processes and kernel methods, it is
sometimes possible to utilize variable selection approaches used for kernel methods also with Gaussian processes.
For example, \citet{crawford2018bayesian,crawford2018variable} derive an analog for the effect size of each
input variable for nonparametric methods, and show that it generalizes also to Gaussian process models. They then 
assess the importance of variable $j$ using Kullback-Leibler divergence between the marginal distribution of the
rest of the variables to their conditional distribution when variable $j$ is
set to zero.

The main contributions of this paper are summarized as follows.
We present two novel input variable selection methods for Gaussian process models, which
directly assess the predictive relevance
of the variables via sensitivity analysis.
Both methods utilize the posterior of the full model (i.e.\ one that includes all the input variables) near the
training points to estimate the predictive relevance of the variables.
We also demonstrate why certain properties of automatic relevance determination make it unsuitable for
variable selection and show that the proposed methods are not affected by these weaknesses.
Our empirical evaluations indicate that the proposed methods lead to improved variable selection
performance in terms of predictive performance, and
generate the relevance ranking more consistently
between different training data sets.
We also demonstrate how the pointwise estimates of the methods
can be useful for assessing local predictive relevance beyond
the global, average relevance of each variable.
The methods serve as practical alternatives to automatic relevance determination
without increasing the computational cost as much as many alternative variable selection methods in the literature.

\section{BACKGROUND}

This section shortly reviews Gaussian processes
and automatic relevance determination in this context, as well as discusses
the problems associated with variable selection via ARD.

\subsection{Gaussian Process Models} \label{sec:GP}

Gaussian processes (GPs) are nonparametric models that define a prior distribution directly in the
space of latent functions $f (\mathbf{x})$, where $\mathbf{x}$ is
a $p$-dimensional input vector.
The form and smoothness of the functions generated by a GP are determined by
its covariance function $k (\mathbf{x},\mathbf{x'})$, which defines
the covariance between the latent function values at the input points $\mathbf{x}$ and $\mathbf{x'}$.
The prior is typically assumed to have zero mean:
\begin{equation*}
p ( f (\mathbf{X})) = p ( \mathbf{f}) = \mathcal{N} (\mathbf{f} \, | \, 0 , \mathbf{K}) ,
\end{equation*}
where $\mathbf{K}$ is the covariance matrix between the latent function values
at the training inputs $\mathbf{X} = (\mathbf{x}^{(1)}, \ldots , \mathbf{x}^{(n)})$
such that
$\mathbf{K}_{ij} = k (\mathbf{x}^{(i)},\mathbf{x}^{(j)})$.

In regression problems with Gaussian observation models, the GP posterior distribution is analytically tractable
for both the latent values and noisy observations
by conditioning the joint normal distribution
of training and test outputs on the observed data.
While other observation models
do not have analytically tractable solutions, numerous
approximations have been developed for inference with different likelihoods in both regression and classification~\citep{williams1998bayesian,minka2001family,vanhatalo2009gaussian}.

\subsection{Automatic Relevance Determination}

A widely used covariance function in Gaussian process inference is the squared exponential (SE) with separate length-scale parameters~$l_i$ for each of the $p$ input dimensions
\begin{equation} \label{eq:ARD-SE}
k_{\mathrm{SE}} (\mathbf{x},\mathbf{x'}) = \sigma_f^2 \, \mathrm{exp} \left( - \frac{1}{2} \sum_{i = 1}^p \frac{(x_i - x_i')^2}{l_i^2} \right ).
\end{equation}
While the common hyperparameter~$\sigma_f$ determines the overall variability,
the separate length-scale parameters~$l_i$
allow
the functions to vary at different scales
along different variables.

In some contexts, automatic relevance determination (ARD) simply means using
the covariance function~(\ref{eq:ARD-SE}) instead of one with a single length-scale parameter~$l$.
Often, however, the term is used for a more specific meaning, namely
for inferring the predictive relevance of each variable
from the inverse of its length-scale parameter.
This intuition is based on the fact that an infinitely
large length-scale means no correlation between the latent function
values in that dimension.
However, in practice they will never be infinite, and
inferring irrelevance from a large length-scale is problematic for two reasons.
First, the length-scale parameters alone are not well identified, but only the ratios
of~$l_i$ and~$\sigma_f$~\citep{zhang2004inconsistent}, which increases variance of the relevance
measure.
Second, ARD systematically overestimates the predictive relevance of nonlinear
variables relative to linear variables of equal relevance in the squared error
sense~\citep{piironen2016projection}.

\section{PREDICTING FEATURE RELEVANCES VIA SENSITIVITY ANALYSIS}

This section describes the two proposed variable selection methods
that analyze the sensitivity of the posterior GP at the training data locations.
We will first present
the outline of both methods and their properties, and
then discuss their computational complexity.

\subsection{Kullback-Leibler Divergence as a Measure of Predictive Relevance} \label{sec:KL}

The Kullback-Leibler divergence (KLD) is a widely used measure of dissimilarity between two probability
distributions~\citep{kullback1951information}.
In this section, we present a method for assessing the predictive relevance
of input variables via sensitivity analysis of the posterior predictive distribution.
When moving an input with respect to a single variable, a large difference in KLD between predictive distributions indicates that the variable has a high predictive relevance.
In our method, the predictive distributions are compared at the training points and
points that are moved with respect to one variable.
The KLD is a favourable measure for predictive relevance, because it takes into
account changes in both the predictive mean and uncertainty.
As the KLD is applicable to arbitrary distributions with compatible support, this procedure
is not limited to any specific likelihood or model.

By relating to the total-variation distance
of Pinsker's inequality, it is reasonable to utilize the Kullback-Leibler divergence
from density~$p$~to~$q$, $\mathcal{D}_{\text{KL}} (p \, || \, q)$,
as a measure of distance in the form~\citep{simpson2017penalising}
\begin{equation*}
d (p \, || \, q) = \sqrt{2 \, \mathcal{D}_{\text{KL}} (p \, || \, q)} .
\end{equation*}
Using the square root also allows linear approximation of infinitesimal changes in the predictive distribution via perturbations in the input variables.
Applying this to the posterior predictive distribution at a training point~$i$, $p(y_* | \mathbf{x}^{(i)}, \mathbf{y})$, and
a point that is perturbed by amount~$\Delta$ with respect to variable~$j$, $p(y_* | \mathbf{x}^{(i)} + \Delta_j, \mathbf{y})$, we use
the following measure of predictive relevance
\begin{equation} \label{eq:KLrelev}
r ( i, j,  \Delta) = \frac{ \, d ( p(y_* | \mathbf{x}^{(i)}, \mathbf{y}) \, || \, p(y_* | \mathbf{x}^{(i)} + \Delta_j, \mathbf{y}) ) }{\Delta}, 
\end{equation}
where $\Delta_j$ is a vector of zeroes with $\Delta$ on the~$j$'th entry.
Averaging this measure over all of the training points~$i$ yields a 
relevance estimate for the $j$'th variable
\begin{align*}
  \text{KL}_j = \frac{1}{n}\sum_{i=1}^n r ( i, j,  \Delta).
\end{align*}
Using this estimate, we can rank the input variables by relevance
and  a desired number of them can be selected.
Henceforth, we will refer to the presented method as the KL method.

For Gaussian observation models, the relevance measure defined in equation~\eqref{eq:KLrelev} is related to the partial derivative of the mean of the latent function with respect to the variable~$j$. 
In this case, if the latent variance is constant, taking the limit $\Delta \rightarrow 0$ simplifies the measure in equation~\eqref{eq:KLrelev} to the form
\begin{equation*}
\lim_{\Delta \to 0} r ( i, j,  \Delta) = ( \text{Var} [ y_* | \mathbf{x}^{(i)}, \mathbf{y}     ] )^{-1/2}   \frac{\partial}{\partial x_j} \mathbb{E} [ y_* | \mathbf{x}^{(i)}, \mathbf{y} ],
\end{equation*}
where $\mathbb{E} [ y_* | \mathbf{x}^{(i)}, \mathbf{y} ]$~and~$\text{Var} [  y_* | \mathbf{x}^{(i)}, \mathbf{y}     ]$ are the mean and variance of the posterior predictive distribution $p(y_* | \mathbf{x}^{(i)}, \mathbf{y})$, respectively.
Hence, in the special case of a Gaussian likelihood, the proposed measure can be interpreted as a partial derivative weighted by the predictive uncertainty. 
The method thus has a connection to methods that rank variables via partial derivatives, see e.g.~\citep{hardle1989investigating,ruck1990feature,lal2006embedded,liu2018model}.

The choice of the perturbation distance $\Delta$ has to be reasonable
with respect to the given data set. According to our empirical evaluations, the proposed method is insensitive to the size of the perturbation. The results of this
paper are computed with $\Delta = 10^{-4}$
when the inputs were normalized to zero mean and unit standard deviation, and
no noticeable differences
were observed when~$\Delta$ was varied for two orders of magnitude above and below this value.
However, very small values
should be avoided because of potential numerical instability. For more details, see Figure~\ref{fig:deltatest} in the supplementary material.

\subsection{Variance of the Posterior Latent Mean}

In this section, we present a method for ranking input variables
based on the variability of the GP latent mean
in the direction of each variable.
When the value of a single input variable is changed,
large variability in the latent mean indicates that the variable is relevant for predicting
the target variable.
In contrast to the KL method, this method thus considers only the latent mean, but
examines it throughout the conditional distribution of each variable at the training point and not just the immediate vicinity
of the point.
We thus ignore the uncertainty
of the predictions but utilize information from a larger area of the input space.
Another benefit of this is that computing the predictive mean of a Gaussian process is computationally cheaper
than predicting the marginal variance.

In order to estimate the variance of the mean of the latent function, we will approximate the distribution of the input variables.
Under the assumption that the input data has finite first and second moments, we can do this by computing the sample mean
$\boldsymbol{\mu}$ and sample covariance $ \boldsymbol{\Sigma}$ from the $n$ training inputs $\mathbf{X} = (\mathbf{x}^{(1)}, \ldots , \mathbf{x}^{(n)})$.
At any given point, the conditional distribution of variable $j$ at training point $i$, $p ( x_j \, | \, \mathbf{x}^{(i)}_{-j} )$, can then be estimated using the conditioning rule of the multivariate Gaussian:
\begin{equation} \label{eq:VARmethodcond}
\begin{split}
x_j \, | \, \mathbf{x}_{-j} & \sim \mathcal{N} ( m_j , s_j^2 ) ,  \\
m_j & = \mu_j + \boldsymbol{\sigma}_{j,-j} \boldsymbol{\Sigma}_{-j,-j}^{-1} (\mathbf{x}_{-j} - \boldsymbol{\mu}_{-j}) , \\
s_j^2 & = \sigma_{j,j} - \boldsymbol{\sigma}_{j,-j} \boldsymbol{\Sigma}_{-j,-j}^{-1} \boldsymbol{\sigma}_{-j,j} .
\end{split}
\end{equation}
Here, the subscript~$j$ refers to selecting the row or column~$j$ from~$\boldsymbol{\mu}$
or~$\boldsymbol{\Sigma}$, whereas the subscript~$-j$ refers to
excluding them.

The variance of the posterior mean ${\overline{f}_j^{(i)} = \mathbb{E} [ f (x_j) \, | \, \mathbf{x}_{-j}^{(i)} ]}$ along the $j$'th dimension is then given by integrating over the conditional distribution  $p ( x_j \, | \, \mathbf{x}^{(i)}_{-j} )$
\begin{align*}
\mathrm{Var} [\overline{f}_j^{(i)}]  = & \int (\overline{f}_j^{(i)})^2 (x_j)  \, \mathcal{N} ( x_j \, | \, m_j , s_j^2 ) \, \mathrm{d} x_j \\
 - & \left (  \int \overline{f}_j^{(i)} (x_j) \, \mathcal{N} ( x_j \, | \, m_j , s_j^2 ) \, \mathrm{d} x_j \right )^2 .
\end{align*}
With a change of variables $k = (x_j - m_j)/(\sqrt{2} s_j)$, the variance
takes a simpler form that can be numerically approximated with the Gauss-Hermite quadrature.
\begin{align*}
\mathrm{Var} [\overline{f}_j^{(i)}]  = & \int (\overline{f}_j^{(i)})^2 (\sqrt{2} s_j k + m_j) \frac{e^{-k^2}}{\sqrt{\pi}} \, \mathrm{d} k \\
 - & \left ( \int \overline{f}_j^{(i)} (\sqrt{2} s_j k + m_j) \frac{e^{-k^2}}{\sqrt{\pi}}   \, \mathrm{d} k \right )^2  \\
  \approx & \, \pi^{-1/2} \sum_{{n_k}=1}^{N_k} w_i \, (\overline{f}_j^{(i)})^2 (\sqrt{2} s_j k_i + m_j) \\
   - & \, \pi^{-1} \left ( \sum_{{n_k}=1}^{N_k} w_i \, \overline{f}_j^{(i)} (\sqrt{2} s_j k_i + m_j) \right )^2 ,
\end{align*}
where $N_k$ is the number of weights $w_i$ and evaluation points $k_i$ of
the Gauss-Hermite quadrature approximation.
The $k_i$ are given by the roots of the physicists' version of the Hermite polynomial $H_{N_k} (k)$ and
the weights are
\begin{equation}
w_i = \frac{2^{N_k - 1} N_k ! \sqrt{\pi}}{N_k^2 [H_{N_{k-1}} (k_i)]^2} .
\end{equation}
The above procedure is repeated with
all of the conditional distributions $p ( x_j \, | \, \mathbf{x}^{(i)}_{-j} )$ from the~$n$ training points and the average is computed
\begin{align}
  \text{VAR}_j = \frac{1}{n}\sum_{i=1}^n \mathrm{Var} [\overline{f}^{(i)}_j].
\end{align}
The average is then used as an estimate of the predictive relevance of variable~$j$.
Henceforth, we will refer to this method as the VAR method.

In this paper, we will only consider data sets where the number of data points $n$
is greater than the number of input dimensions $p$.
In the absence of linearly dependent components in the inputs,
the resulting sample covariance matrix $ \boldsymbol{\Sigma}$ will be positive definite and its
inverse can be computed using the Cholesky decomposition.
In order to increase the numerical
stability of the decomposition, a small diagonal term is added to
ill-conditioned sample covariance matrices.
By using more shrinkage when estimating the covariance matrix, the VAR method could be used also
with data sets where $n < p$.

Relating to the KL method, the advantage of the VAR method is that modelling the distribution of the inputs
allows us to examine the GP posterior for out-of-sample behavior in a larger area of the input space than just at the training data locations.
On the other hand, estimating the input distribution is a task itself, which may increase the variance of the resulting
relevance estimate
when data are scarce.

\subsection{Computational Complexity} \label{sec:complexity}

Exact inference with Gaussian processes has complexity $\mathcal{O} (n^3)$ for
a data set with $n$ observations, which hinders their applicability especially in large data sets.
Once a full GP model is fitted, ranking variables
using ARD requires no additional computations.
By a projection approach~\citep{piironen2016projection}, the variables
can possibly be ranked more accurately, but the drawback is that the model space exploration to find the submodels
increases the complexity 
to $\mathcal{O} (p^2 n^3)$, where $p$ is the number of input variables.

The complexity of Gaussian process inference arises from the unavoidable matrix inversion.
However, the same inverse can be used for making an arbitrary number of
predictions at new test points,
and the cost of predicting the GP mean and variance at a single test point are
$\mathcal{O} (n)$ and $\mathcal{O} (n^2)$, respectively.
Both of the proposed methods in this paper utilize a constant number of predictions
for each of the~$n$ data points and~$p$ input variables.
As the VAR method does not require the predictive variance, its computational complexity
is $\mathcal{O} (p \cdot n \cdot n) = \mathcal{O} (p n^2)$, whereas the KL method has complexity $\mathcal{O} (p n^3)$.
For both methods, using a sparse GP approximation with $m < n$ inducing
points can reduce the cost of predictions and reduce the complexity of the proposed methods to $\mathcal{O} (p n m)$~and~$\mathcal{O} (p n m^2)$, respectively for the VAR and KL methods~\citep{bui2017unifying}.
Alternatively, one may reduce computational cost by using only a subset of training points to estimate the predictive
relevances of variables.

In addition to the complexity due to predictions, the VAR method 
requires inverting the submatrix $\boldsymbol{\Sigma}_{-j,-j}$ of the
sample covariance matrix of the inputs, for
each of the $p$ variables.
Taking advantage of the positive definiteness of the full covariance matrix,
the Cholesky decomposition of it, $\mathcal{O} (p^3)$ in complexity, needs
to be computed only once per training set.
Then the Cholesky decomposition for each submatrix $\boldsymbol{\Sigma}_{-j,-j}$
is obtained with a rank one update from the full covariance matrix, resulting
in $p$ rank one updates of complexity $\mathcal{O} (p^2)$.
Thus, the full complexity of the variance method
is $\mathcal{O} (p n^2 + p^3)$.
Because we are considering only the case $p < n$, the effective
complexity is still $\mathcal{O} (p n^2)$.
The details of the rank one update are described in the supplementary material.

\section{EXPERIMENTS}

This section will present two toy examples that illustrate how the proposed methods
are able to assess the predictive relevance between linear and nonlinear variables
more accurately compared to automatic relevance determination.
The section will also present variable selection results
in regression and classification tasks on real data sets. In all experiments, the model of choice
is a GP model with the ARD covariance function in equation~(\ref{eq:ARD-SE}).
We want to emphasize, that our intent is not to criticize the use of this kernel in general, but
to show that it is problematic to use it for assessing the relevance of input variables.

\subsection{Toy Examples} \label{sec:toymodel}

\begin{figure}[tb]
  \centering
    \includegraphics[width=0.5\textwidth]{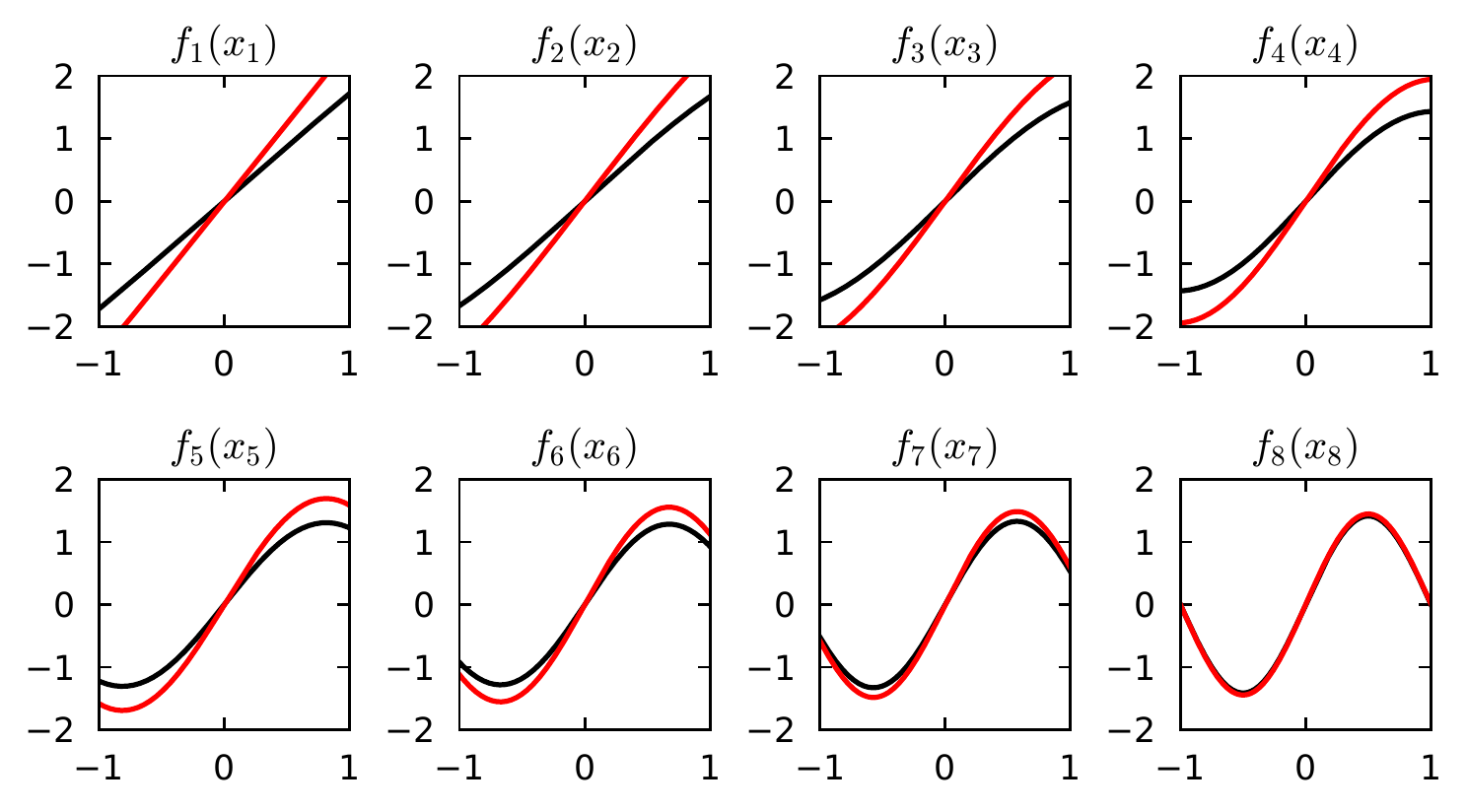} 
\caption{Latent functions $f_j (x_j), j = 1, \ldots , 8$
of the two toy examples. Black represents the latent functions with uniform inputs and
red represents normally distributed inputs, each function scaled to unit variance according to
its corresponding input distribution.} \label{fig:toyplot}
\end{figure}

In the first experiment we consider a toy example, where
the target variable is constructed as a sum of
eight independent and additive variables whose responses have varying degrees of nonlinearity.
We generate the target variable $y$ based on the inputs as follows:
\begin{equation} \label{eq:toymodel}
\begin{split}
y & = f_1 (x_1) + \ldots + f_8 (x_8) + \varepsilon , \\
\varepsilon & \sim \mathcal{N} (0,0.3^2) , \\
f_j (x_j) & = A_j \sin{ ( \phi_j x_j ) } ,
\end{split}
\end{equation}
where the angular frequencies $\phi_j$ are equally spaced between $\pi / 10$ and $\pi$, and
the scaling factors $A_j$ are such that the variance of each $f_j (x_j)$ is one.
We consider two separate mechanisms for generating the input data
so that either $x_j \sim \mathrm{U} (-1,1) $ or $ x_j \sim \mathcal{N} (0,0.4^2)$.
The functions $f_j$ are presented in Figure~\ref{fig:toyplot} for
uniformly distributed inputs (black) and normally distributed inputs (red).

\begin{figure}[tp]
  \centering
    \includegraphics[width=0.5\textwidth]{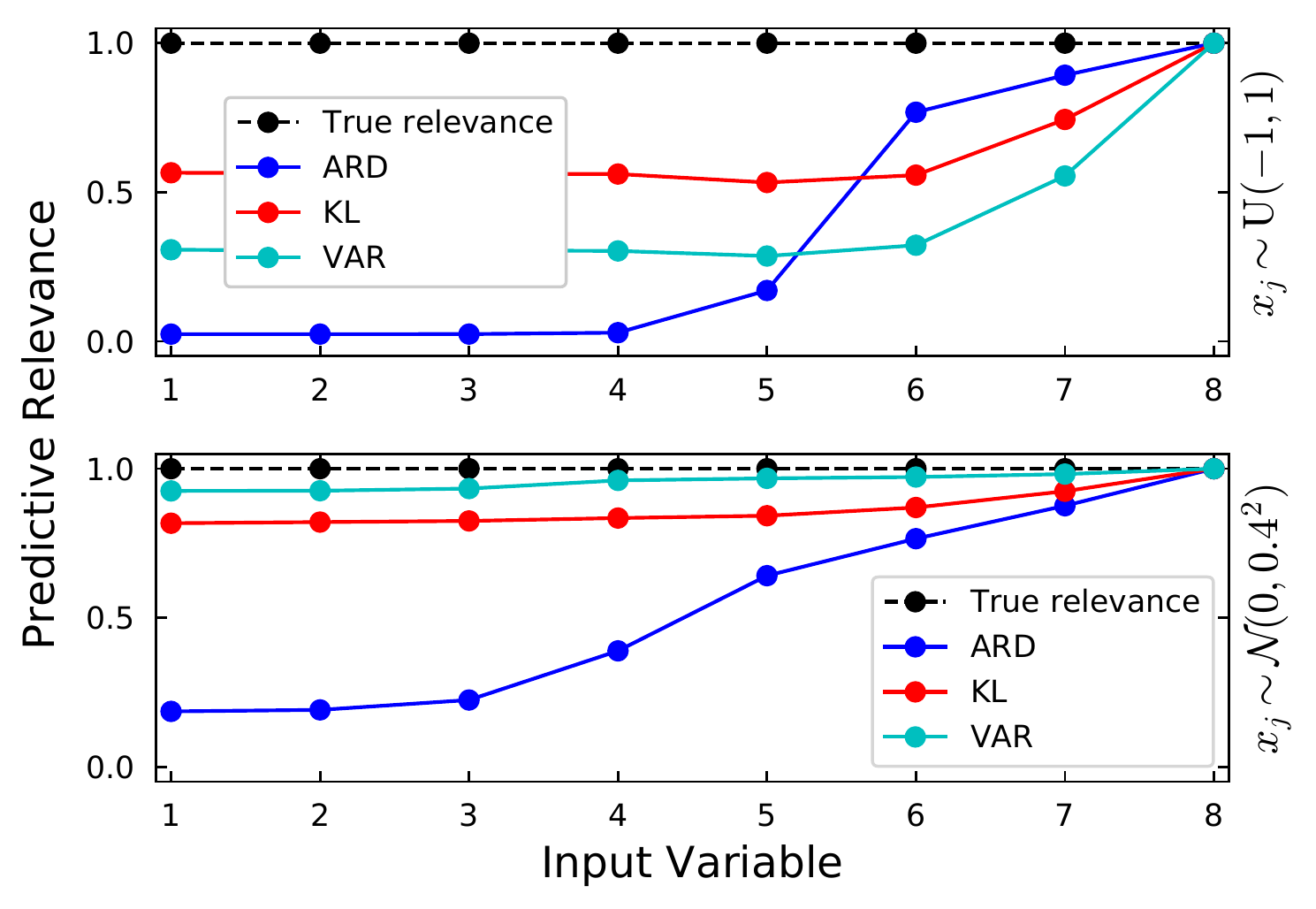}
\caption{Relevance estimates for eight variables in the two toy examples in equation~(\ref{eq:toymodel}) with uniformly
distributed inputs (top) and normally distributed inputs (bottom).
The estimates are computed
with ARD (blue), the KL method (red), and the VAR method (cyan).
The results are averaged over 200 data realizations and scaled so that
the most relevant variable has a relevance of one. Error bars representing $95\%$ confidence intervals are indistinguishable.
} \label{fig:toyorders}
\end{figure}

For both toy examples, we sampled 300 training points
and
constructed a Gaussian process model with a covariance function
being the squared exponential in equation~(\ref{eq:ARD-SE}) with an added constant term.
Using the full model with hyperparameters optimized
to the maximum of marginal likelihood, we calculated the relevance of each variable either directly via ARD using the inverse length-scale,
or by averaging the KL and VAR relevance estimates from each training point.
The averaged results of 200 random data sets are presented in Figure~\ref{fig:toyorders}
for the two examples with inputs distributed uniformly (top) and normally (bottom). Input 1
is the most linear one and input 8 is the most nonlinear.

\begin{figure*}[tpb]
  \centering
    \includegraphics[width=0.99\textwidth]{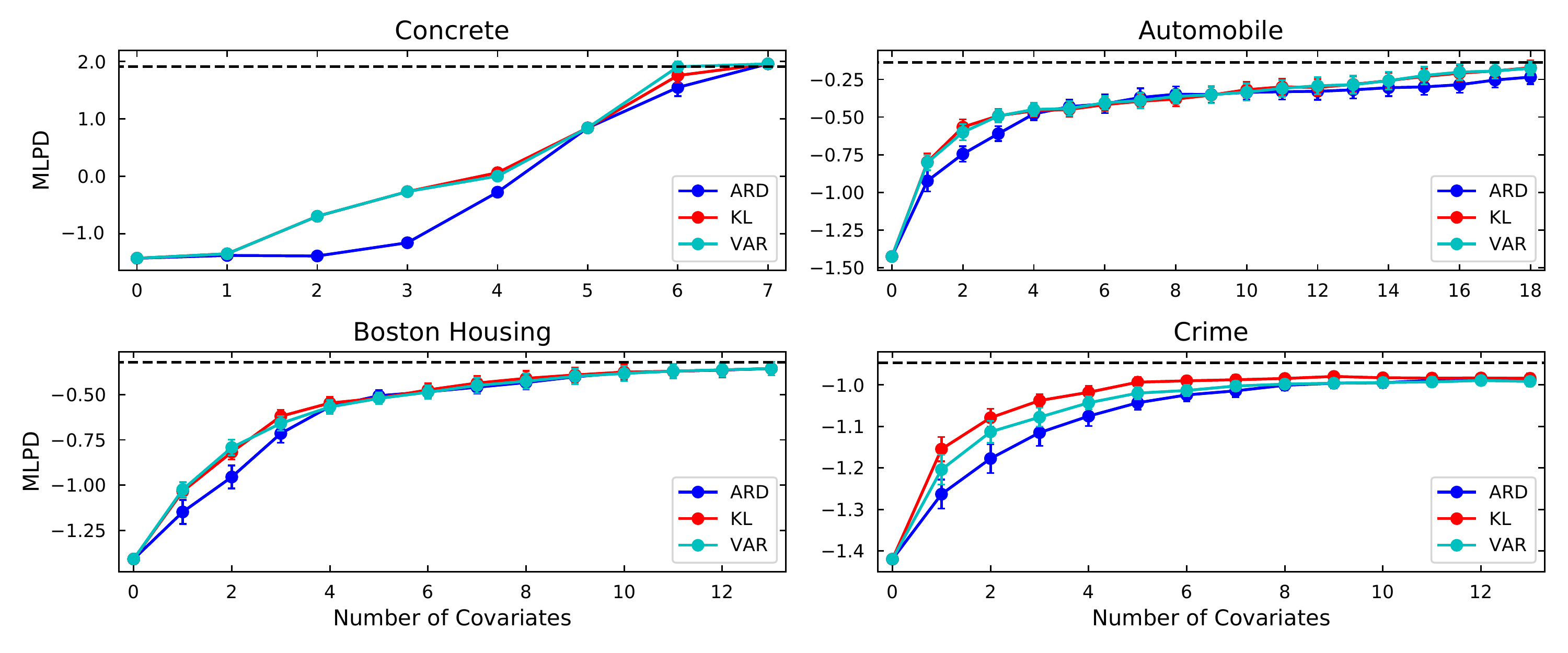}
\caption{Mean log predictive densities (MLPDs) of the test sets with $95 \%$ confidence
intervals for submodels as a function of variables included in the submodel.
Blue depicts variables sorted using ARD, red and cyan depict the KL and VAR methods, respectively. The dashed horizontal line depicts the
MLPD of the full model with hyperparameters sampled using the Hamiltonian Monte Carlo algorithm.} \label{fig:allmlpds}
\end{figure*}

Figure~\ref{fig:toyorders} demonstrates that in the toy example with uniform inputs, all three methods
prefer nonlinear inputs over linear inputs.
However, the preference in our methods is not as severe as with ARD,
which assigns relevance values close to zero for half of the variables.
The bottom figure, representing the toy example with Gaussian distributed inputs, shows that
our methods produce almost equal relevance values for all eight variables.
Overall, our methods are notably better than ARD in identifying the true relevances
of the variables despite the varying degrees of nonlinearity.
To ensure that the above results hold even if there are irrelevant variables
in the data, we repeated the experiment with the addition of totally irrelevant input variables. The results are comparable, and
are shown in Figure~\ref{fig:supp1} in the supplementary material.

\subsection{Real World Data} \label{sec:realworlddata}

In the second experiment, we compared the variable selection performance of the three methods on five benchmark
data sets obtained from the UCI machine learning repository\footnote{\href{https://archive.ics.uci.edu/ml/index.html}{https://archive.ics.uci.edu/ml/index.html}}.
The data sets are summarized
in Table~\ref{table:datasummary}.
The Pima indians data set is a binary classification problem, and the others
are regression tasks.
For each method, we used a Gaussian process model with a Gaussian likelihood and a sum of constant
and squared exponential kernels as a covariance function.
The model was first fitted with all $p$ variables included, and then submodels
with $1$ to $p-1$ of the most relevant variables included were fitted again.
The submodel variables were picked based on
the relevance ranking given by each method.
We performed 50 repetitions, each time splitting the
data into random training and test sets with the number of training points shown in Table~\ref{table:datasummary}.
Both the full model and submodels were
trained on the training set, and the predictive performance of
the submodels was evaluated by computing the mean log predictive densities (MLPDs) using the independent test set.

\begin{table}[bt]
\caption{Summary of real world dataset parameters: number of variables $p$,
data points $n_{\mathrm{tot}}$, and training points used $n$.} \label{table:datasummary}
\begin{center}
\begin{tabular}{llll}
Dataset  & $p$ & $n_{\mathrm{tot}}$ & $n$ \\
\hline \\
Concrete   & 7 & 103 & 80 \\
Boston Housing   & 13 & 506 & 300 \\
Automobile   & 38 & 193 & 150 \\
Crime   & 102 & 1992 & 400 \\
Pima Indians   & 8 & 392 & 300 
\end{tabular}
\end{center}
\end{table}

For the regression tasks, the mean log predictive densities of the submodels on the test sets are presented
in Figure~\ref{fig:allmlpds} as a function of the number of variables included in the submodel.
A plot for each data set contains results when the variables are sorted using ARD (blue), the KL method (red), and
the VAR method (cyan).
Thus, the only difference between the three curves is the choice of variables included in the submodels.
The GP models are fitted by
maximizing the hyperparameter posterior distribution, with a half\nobreakdash-$t$
distribution as the prior
for
the noise and signal magnitudes, and inverse-gamma distribution for the length-scales.
The inverse-gamma was chosen because it has a sharp left tail that penalizes very
small length-scales, but its long right tail
allows the length-scales to become large~\citep{stanmanual}.
The plots for the Automobile and Crime data sets are shown only up to a point where the
predictive performance saturates.
The horizontal line represents the MLPD of the full model on the test sets, which was
computed using 100 Hamiltonian Monte Carlo (HMC) samples from the hyperparameter posterior~\citep{duane1987hybrid}.

\begin{figure}[tb]
  \centering
    \includegraphics[width=0.5\textwidth]{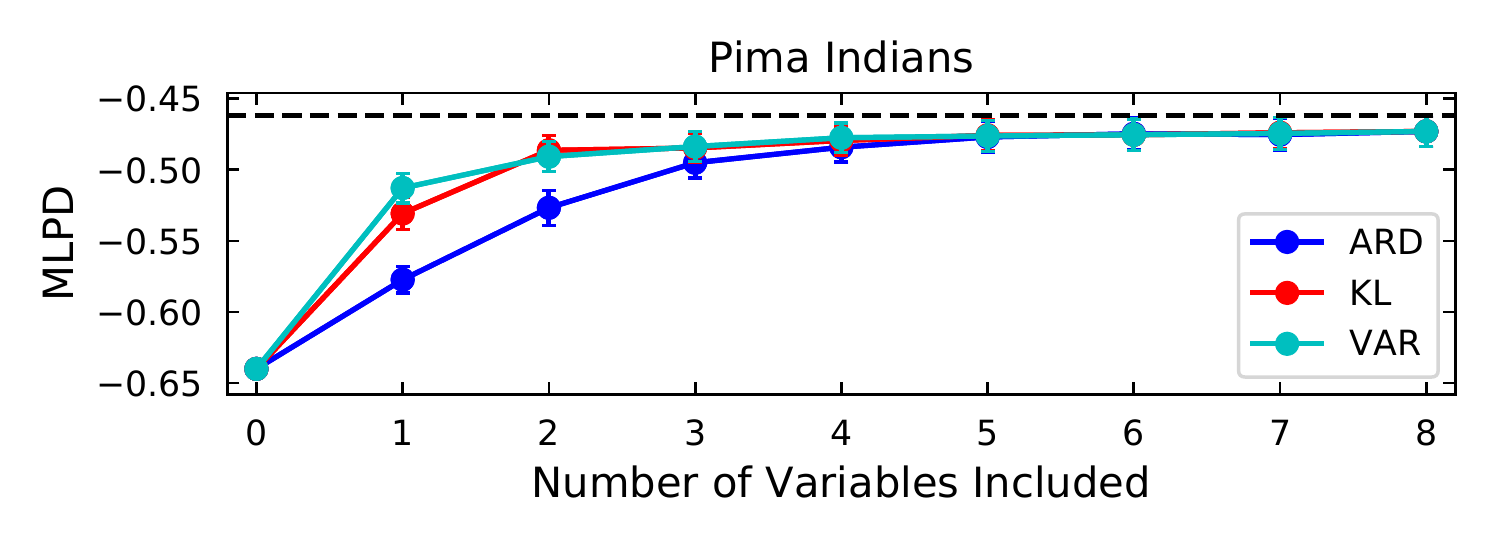} 
\caption{Mean log predictive densities (MLPDs) of the test sets of the Pima indians data set with $95 \%$ confidence
intervals for submodels as a function of the number of variables included in the submodel.
The dashed horizontal line depicts the
MLPD of the full model with hyperparameters sampled using the Hamiltonian Monte Carlo algorithm.} \label{fig:pima}
\end{figure}

\begin{figure*}[tb]
  \centering
    \includegraphics[width=0.99\textwidth]{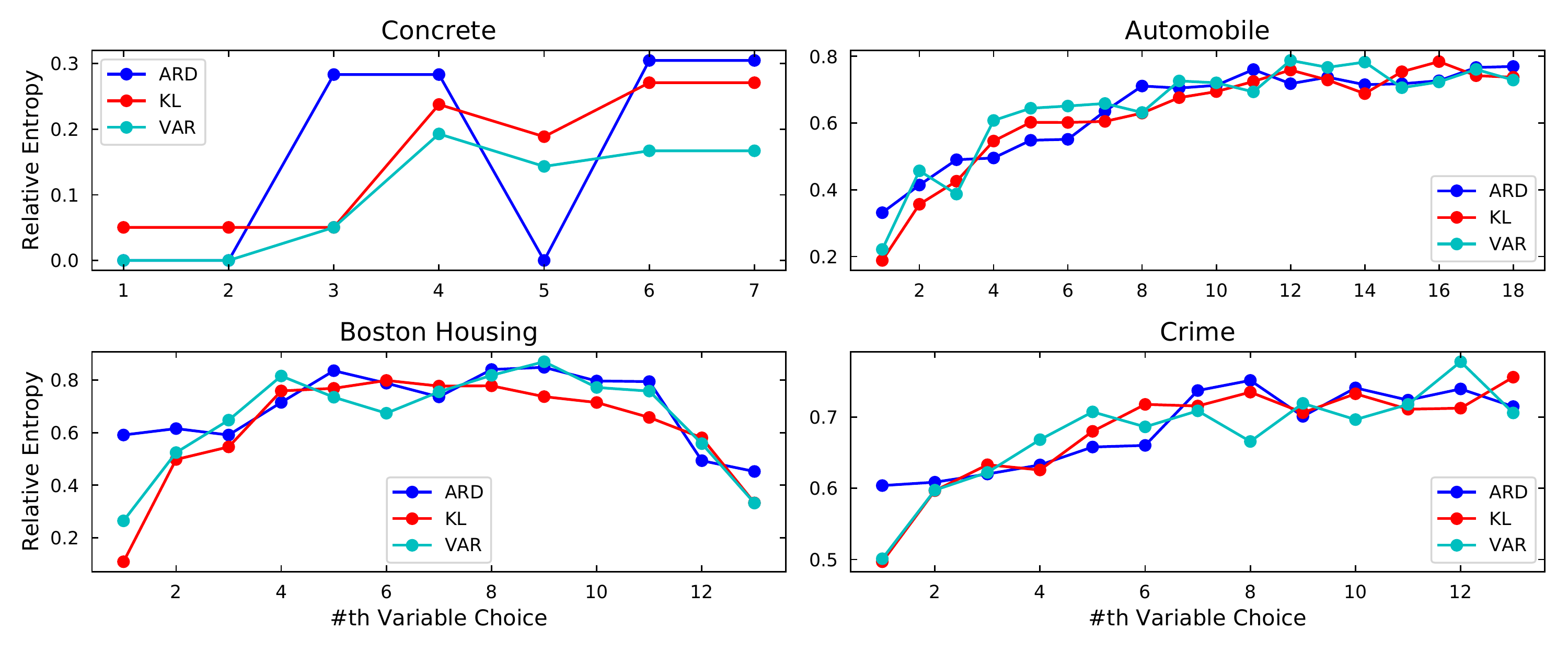}
\caption{Relative entropies that depict the variability between
different training data sets for each consecutive choice of variables to add to the submodel.
Blue depicts variables sorted using ARD, red and cyan depict the KL and VAR methods, respectively.} \label{fig:variabilities}
\end{figure*}

The results show that in all four data sets, both of the proposed methods generate
a better ranking for the variables than ARD does, resulting in submodels with better
predictive performance. The improvement is most distinct in the first three or four variables in all the data sets.
This is because ARD, by definition, picks the most nonlinear variables first, but our methods are able to identify variables that are more relevant for prediction,
albeit more linear.
After the initial improvement, the ranking in the latter variables is never
worse than for ARD.

For the binary classification problem, we used a probit likelihood and the
expectation propagation (EP)~\citep{minka2001family} method to approximate the posterior distribution.
The mean log predictive densities on independent test sets as a function
of the number of variables included in the submodel are presented in Figure~\ref{fig:pima}.
The improvement in variable ranking is very similar to the regression tasks, with largest
improvements in submodels with one to three variables.
Both of the proposed methods are thus able identify variables with good predictive performance in regression as well as classification tasks.

\subsection{Ranking Variability}

The weak identifiability of the length-scale parameters increases the variation
of the ARD relevance estimate. To quantify this,
we studied the variability of each method in determining the
relevance ranking of variables between the 50 random training splits of the four regression data sets.
For each consecutive choice of which variable to add to the submodel, we computed
the entropy, which depicts the variability of the variable choice between different training sets.
If the same variable is chosen in each training set, the resulting entropy is zero, and
more variability leads to higher entropy.
Because the maximum possible entropy depends on the number of variables to choose from, we
divided the entropy values by the maximum possible entropy of each data set.
The maximum entropy corresponds to the case where any of the $p$ variables is chosen with equal probability.
The variability results are presented in Figure~\ref{fig:variabilities}.

Figure~\ref{fig:variabilities} indicates that the ranking variability is correlated with predictive performance
of the submodels, shown in Figure~\ref{fig:allmlpds}.
In the Housing, Automobile, and Crime data sets, ARD has the largest variability
in the first variable choice.
This seems to propagate into improved predictive performance in small submodels
with one to three variables.
On the other hand, in the Concrete data set, ARD has more variability in
the latter variable choices. For example, the better performance of the submodel with six variables is purely the result of
choosing between two variables more consistently, because all three methods always pick the same two variables last, but
ARD is more uncertain about their order of relevance.
A more detailed analysis of the ranking variability is presented in the supplement.

\subsection{Pointwise Relevance Estimates}

In some cases, a variable might have strong predictive relevance in some region, while
being quite irrelevant on average. In some applications, the identification
of such locally relevant variables is important.
Consider a hypothetical regression problem, where the variables represent
measurements to be made on a patient, and the dependent variable represents
the progression of a disease. The information that some measurement
has little relevance on average, but for some patients it is a clear
indication of how far the disease has progressed, may provide essential information
for medical professionals.
In the context of neural networks,
\citet{refenes1999neural} discuss using the maximum of pointwise relevance values
as a useful indicator in financial applications.

\begin{figure}[tb]
  \centering
    \includegraphics[width=0.5\textwidth]{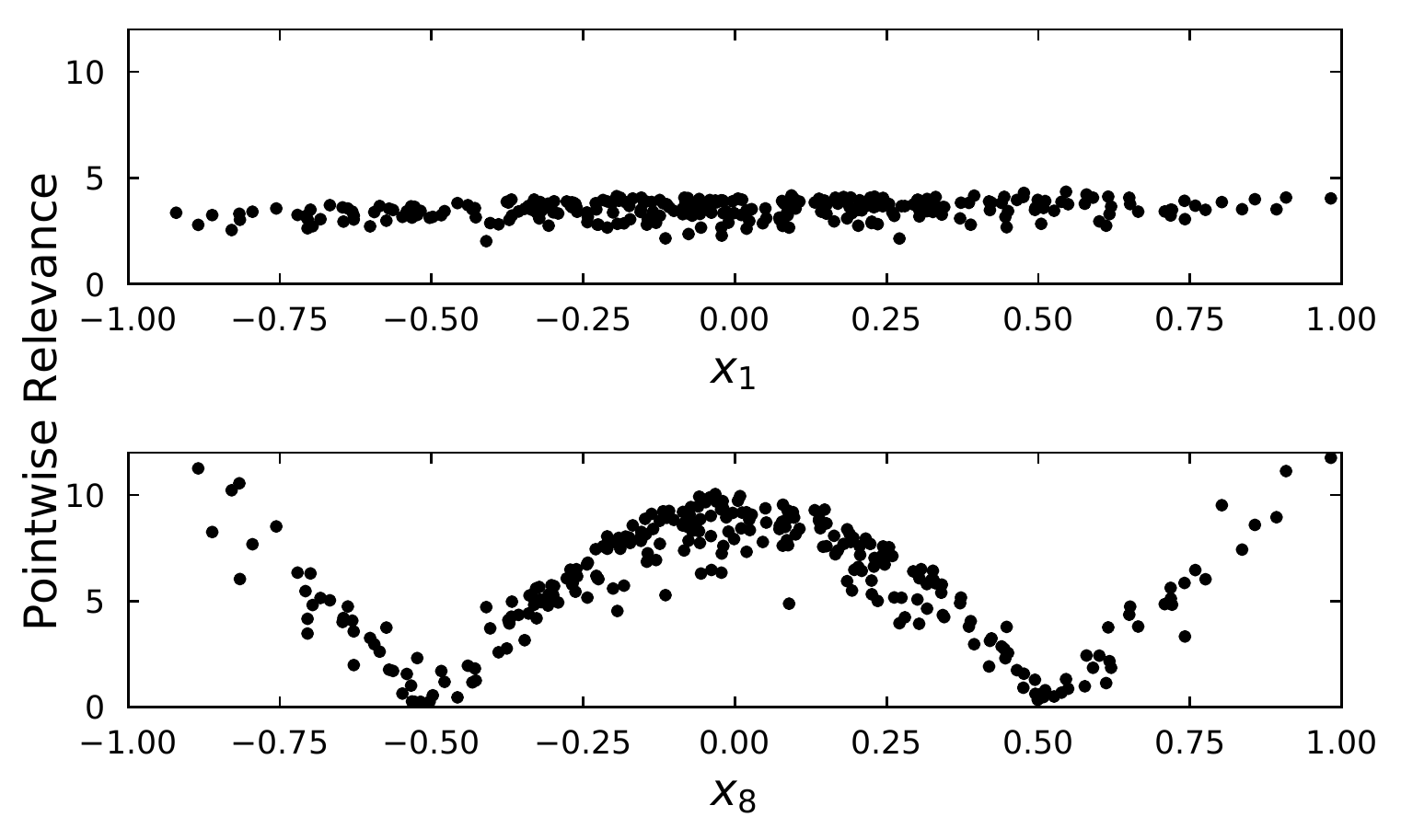}
\caption{Plot of the pointwise KL relevance values
for variables 1 and 8 computed from a sample of 300 points from the toy example in equation~(\ref{eq:toymodel}), where $x_1, x_8 \sim \mathcal{N} (0,0.4^2)$.} \label{fig:localtoy}
\end{figure}

Both of the proposed variable selection methods can be used
to assess the relevance of variables in a specific area of the input space.
To demonstrate this, we
computed the pointwise KL relevance values of the variables 1 and 8
from a sample of 300
training points
from the toy example in equation~(\ref{eq:toymodel}), and the results are presented in Figure~\ref{fig:localtoy}.
As mentioned in Section~\ref{sec:KL}, in Gaussian process regression with a Gaussian likelihood,
the KL relevance value is analogous to the partial derivative of the mean of the latent function divided by the standard deviation
of the posterior predictive distribution. This can be clearly seen by comparing Figure~\ref{fig:localtoy}
to the true latent functions $f_1 (x_1)$ and $f_8 (x_8)$ in Figure~\ref{fig:toyorders}.

The pointwise predictive relevance values presented in Figure~\ref{fig:localtoy}
illustrate one of the novel aspects of the proposed methods.
While automatic relevance determination
outputs only a single number that implicitly represents the relevance of a variable, the KL and VAR methods
can compute the relevance estimate of input variables
at arbitrary points of the input space.
As shown in the experiments of Section~\ref{sec:realworlddata}, averaging the pointwise values at the training points is effective
in assessing the global predictive relevance, and
the ability to consider predictive relevance locally improves the applicability of the methods.

\section{CONCLUSIONS}

This paper has proposed two new methods for ranking variables in Gaussian process models
based on their predictive relevances.
Our experiments on simulated and real world data sets
indicate that the methods produce an improved variable relevance ranking
compared to the commonly used automatic relevance determination via length-scale parameters.
Regarding the predictive performance, although the methods were better than
ARD, even better results could be obtained by other means, such as the predictive projection method, but
at the expense of a much higher computational cost.
Additionally, our methods were shown
to generate the relevance ranking for variables with less variation compared to ARD, which is an important
result in terms of interpretability of the chosen submodels.
We also showed how one of the methods is connected to relevance estimation via derivatives, which
encourages further research in this direction.

The methods proposed here require computing relevance values for each variable in each point of the
training data, thus increasing the computational cost compared
to automatic relevance determination.
However, this cost is by no means prohibitive compared to the Gaussian process inference, which
is computationally expensive in itself.
Additionally, the methods are simpler and computationally cheaper than most
alternative methods proposed in the literature.
We thus discourage interpreting the length-scale of a particular dimension as a measure of predictive relevance, and advise
using a more appropriate method for variable selection and relevance assessment.

Python implementations for the methods discussed in the paper are freely available at \href{https://www.github.com/topipa/gp-varsel-kl-var}{https://www.github.com/topipa/gp-varsel-kl-var}.

\bibliography{klpred}

\clearpage
\section*{SUPPLEMENTARY MATERIAL}

\subsection*{KL Method Relevance Measure Equations}

\subsubsection*{Gaussian Observation Model}

For a Gaussian observation model, the predictive distribution of a Gaussian process model at a single test point is a univariate normal distribution.
Let us denote the mean and variance of the predictive distribution at
test point $\mathbf{x}^{(i)}$ as
$\mu_i = \text{E} [y_* | \mathbf{x}^{(i)}, \mathbf{y}]$ and $\sigma_i^2 = \text{Var} [y_* | \mathbf{x}^{(i)}, \mathbf{y}]$, respectively. Analogously, denote the mean and variance 
of the predictive distribution at
the perturbed point as
$\mu_{i, \Delta_j} = \text{E} [y_* | \mathbf{x}^{(i)} + \Delta_j, \mathbf{y}]$ and $\sigma_{i, \Delta_j}^2 = \text{Var} [y_* | \mathbf{x}^{(i)} + \Delta_j, \mathbf{y}]$.
The KL divergence between these distributions is
\begin{equation*}
\log{\frac{\sigma_{i, \Delta_j}}{\sigma_{i}}} +  \frac{\sigma_{i}^2 + (\mu_i - \mu_{i, \Delta_j} )^2 }{2 \sigma_{i, \Delta_j}^2}  - \frac{1}{2} .
\end{equation*}
The measure of predictive relevance in equation~(\ref{eq:KLrelev}) is then
\begin{equation*}
\begin{split}
r (i,j,\Delta) & = \\
\frac{\sqrt{2}}{\Delta} & \sqrt{   \log{\frac{\sigma_{i, \Delta_j}}{\sigma_{i}}} +  \frac{\sigma_{i}^2 + (\mu_i - \mu_{i, \Delta_j} )^2 }{2 \sigma_{i, \Delta_j}^2}  - \frac{1}{2} } .
\end{split}
\end{equation*}

\subsubsection*{Binary Classification}

Consider a binary classification problem modelled with a Gaussian process. The predictive distribution at test point $\mathbf{x}^{(i)}$ is
a Bernoulli distribution with success probability denoted as
$\pi_* = p(y_* = 1 | \mathbf{x}^{(i)}, \mathbf{y})$.
The KL divergence between this distribution and the predictive distribution at a perturbed point,
with success probability $\pi_{*,\Delta_j} = p(y_* = 1 | \mathbf{x}^{(i)} + \Delta_j, \mathbf{y})$, is then
\begin{equation*}
 \pi_* \log{\frac{\pi_*}{\pi_{*,\Delta_j}}} + (1 - \pi_*) \log{\frac{1 - \pi_*}{1 - \pi_{*,\Delta_j}}}.
\end{equation*}
The measure of predictive relevance in equation~(\ref{eq:KLrelev}) is then
\begin{equation*}
\begin{split}
r (i,j,\Delta) & = \\
 \frac{\sqrt{2}}{\Delta} & \sqrt{   \pi_* \log{\frac{\pi_*}{\pi_{*,\Delta_j}}} + (1 - \pi_*) \log{\frac{1 - \pi_*}{1 - \pi_{*,\Delta_j}}}} .
\end{split}
\end{equation*}

\subsection*{Sensitivity of the KL Method to perturbation size $\Delta$}

\begin{figure}[htb]
  \centering
    \includegraphics[width=0.5\textwidth]{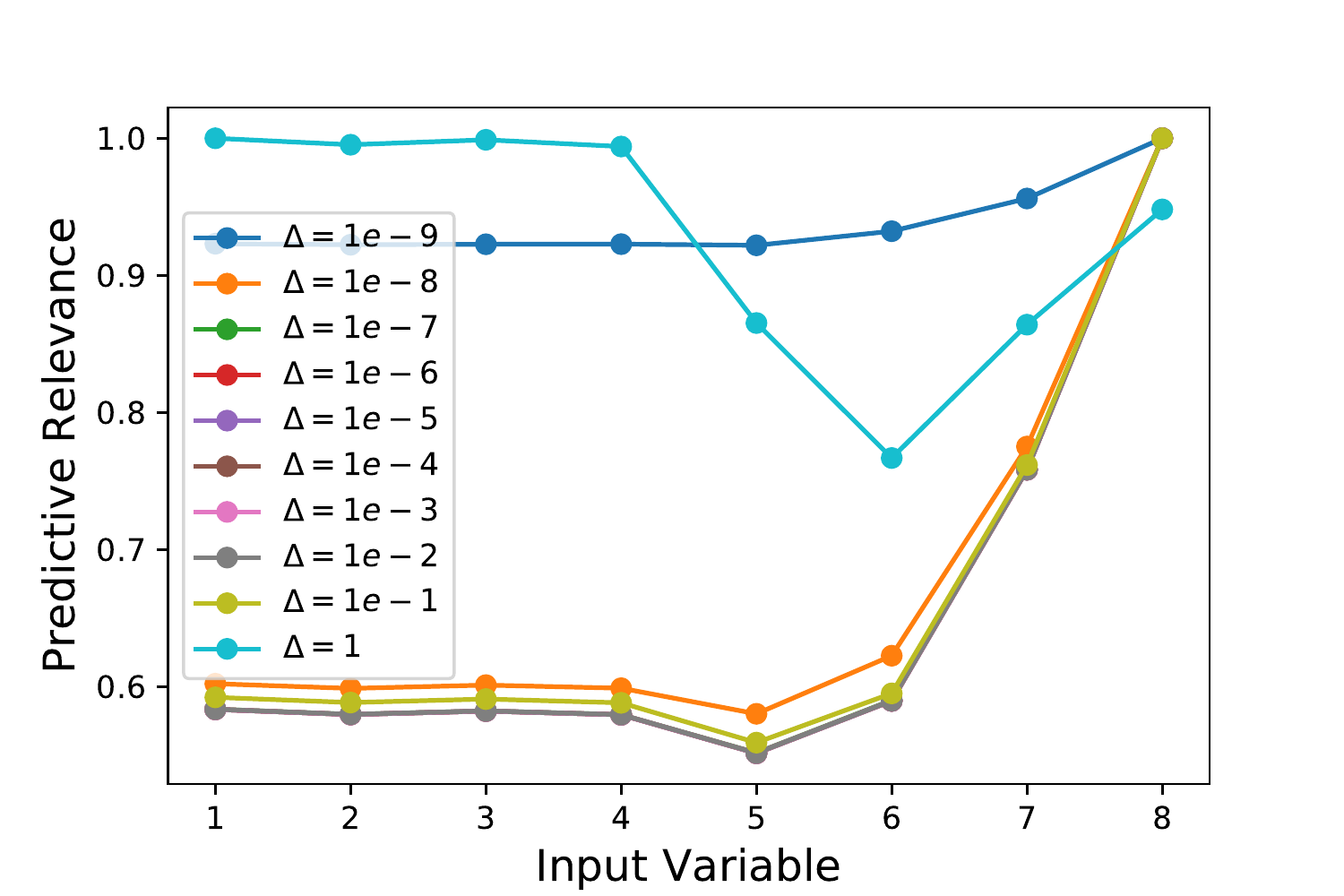} 
\caption{Relevance estimates given by the KL method for eight covariates in the toy example where each variable is equally relevant.
The results are averaged over 50 data realizations and scaled so that
the most relevant covariate has a relevance of one.} \label{fig:deltatest}
\end{figure}

We repeated the toy example from Section~\ref{sec:toymodel} and computed the KL relevance estimates with different values of
the perturbation size $\Delta$. All of the independent input variables have a uniform distribution
$\text{U} (-1,1)$ and thus have a standard deviation of $1/\sqrt{3}$.
Computed relevance estimates of the eight variables averaged from 50 data realizations are plotted in Figure~\ref{fig:deltatest}.
For reasonably small $\Delta$ values the results are identical. 
The results differ only when $\Delta$ is smaller than $10^{-7}$ or larger than $10^{-2}$.
$\Delta = 10^{-4}$ is a safe choice for most purposes unless the inputs have very small length-scale.
In that case, one can make $\Delta$ smaller but should be cautious of numerical errors.

\subsection*{In-depth Look at Ranking Variability}

\begin{figure*}[ptb]
  \centering
    \includegraphics[width=0.85\textwidth]{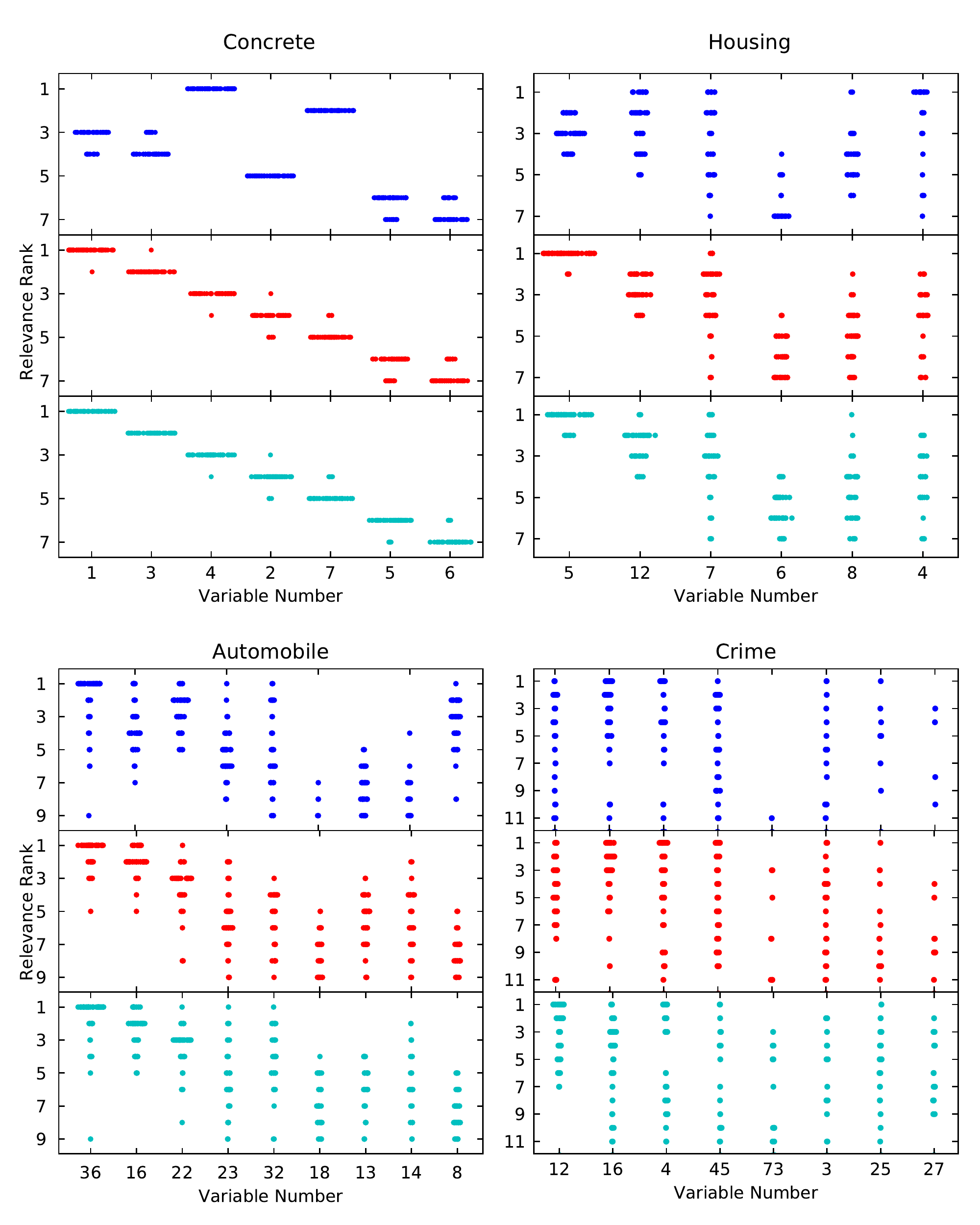}
\caption{A plot representing the variability in relevance ranks between different training sets in the four regression data sets. Blue, red and cyan points represent ARD, KL and VAR ranking methods, respectively.} \label{fig:ordersall}
\end{figure*}

\begin{figure}[htb]
  \centering
    \includegraphics[width=0.48\textwidth]{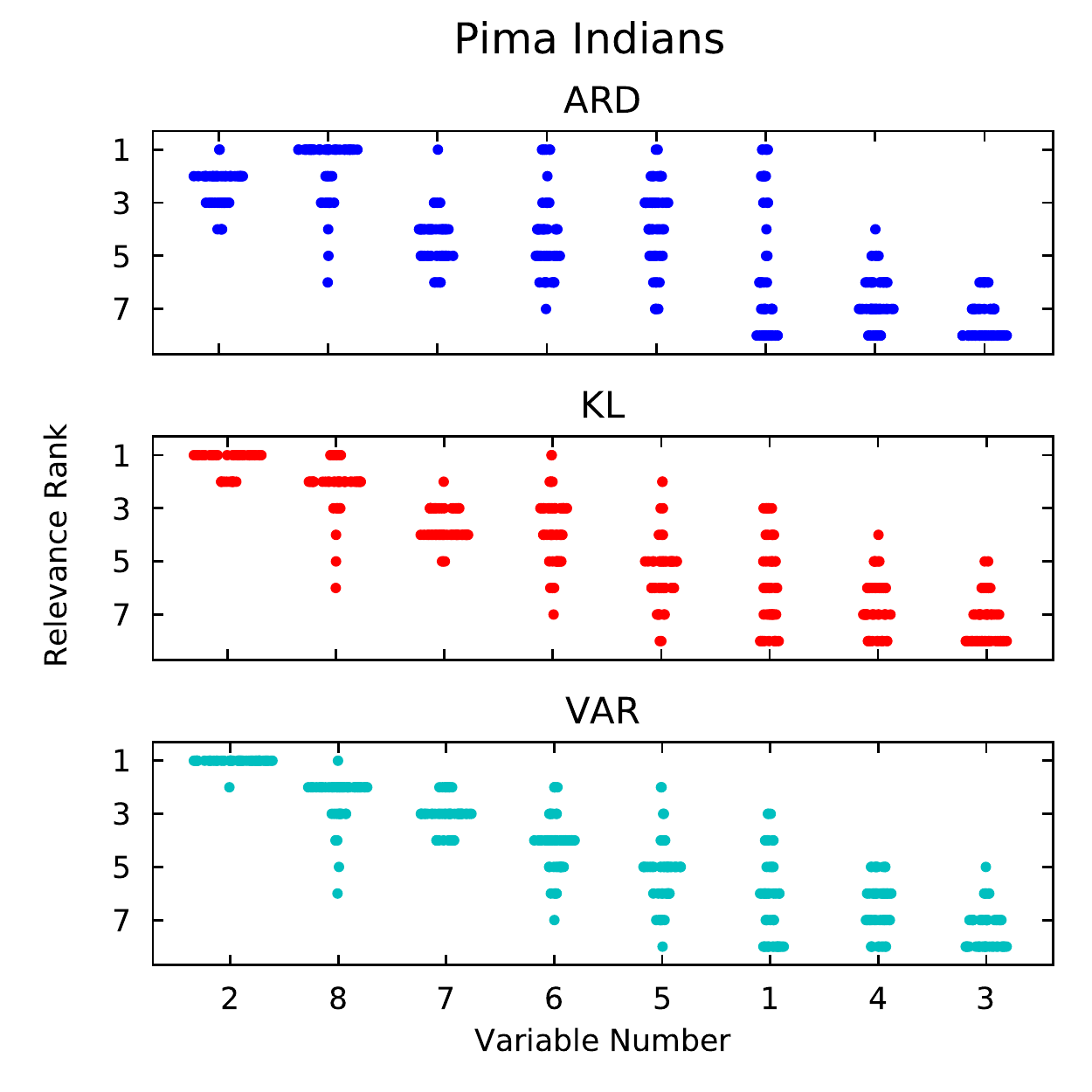}
\caption{A plot representing the variability in relevance ranks between different training sets in the Pima Indians binary classification data set. Blue, red and cyan points represent ARD, KL and VAR ranking methods, respectively.} \label{fig:orderspima}
\end{figure}

To see the effect of ranking variability more clearly, we plotted markers for the variable ranks from each training split
based on 50 training sets from the four regression data sets, and
the results are presented in Figure~\ref{fig:ordersall}.
The markers are jittered horizontally to better illustrate the
number of times each variable was assigned a specific relevance rank.
The
variables are ordered from left to right in terms of highest average relevance given by the KL method.
A similar plot for the Pima Indians data set in shown in Figure~\ref{fig:orderspima}.

For example, the plot of the Concrete data reveals the fact that the improved predictive performance in the chosen submodels 
is not only the result of being able to identify linear but relevant variables, but is also
partly a result of less variation between different training sets. For example, the better performance in the submodel with six variables in Figure~\ref{fig:allmlpds} is strictly the result of
choosing variable 5 more often than variable 6, because all three methods always pick those two last, but
ARD is more unsure about their order.
The Housing data plot shows that while both the KL and VAR methods
pick variable 5 as the most relevant in a majority of training samples, ARD is has more variability,
choosing variables 12, 7, and 4 almost equiprobably.

\subsection*{Toy Example With Irrelevant Variables}

\begin{figure}[h]
  \centering
    \includegraphics[width=0.5\textwidth]{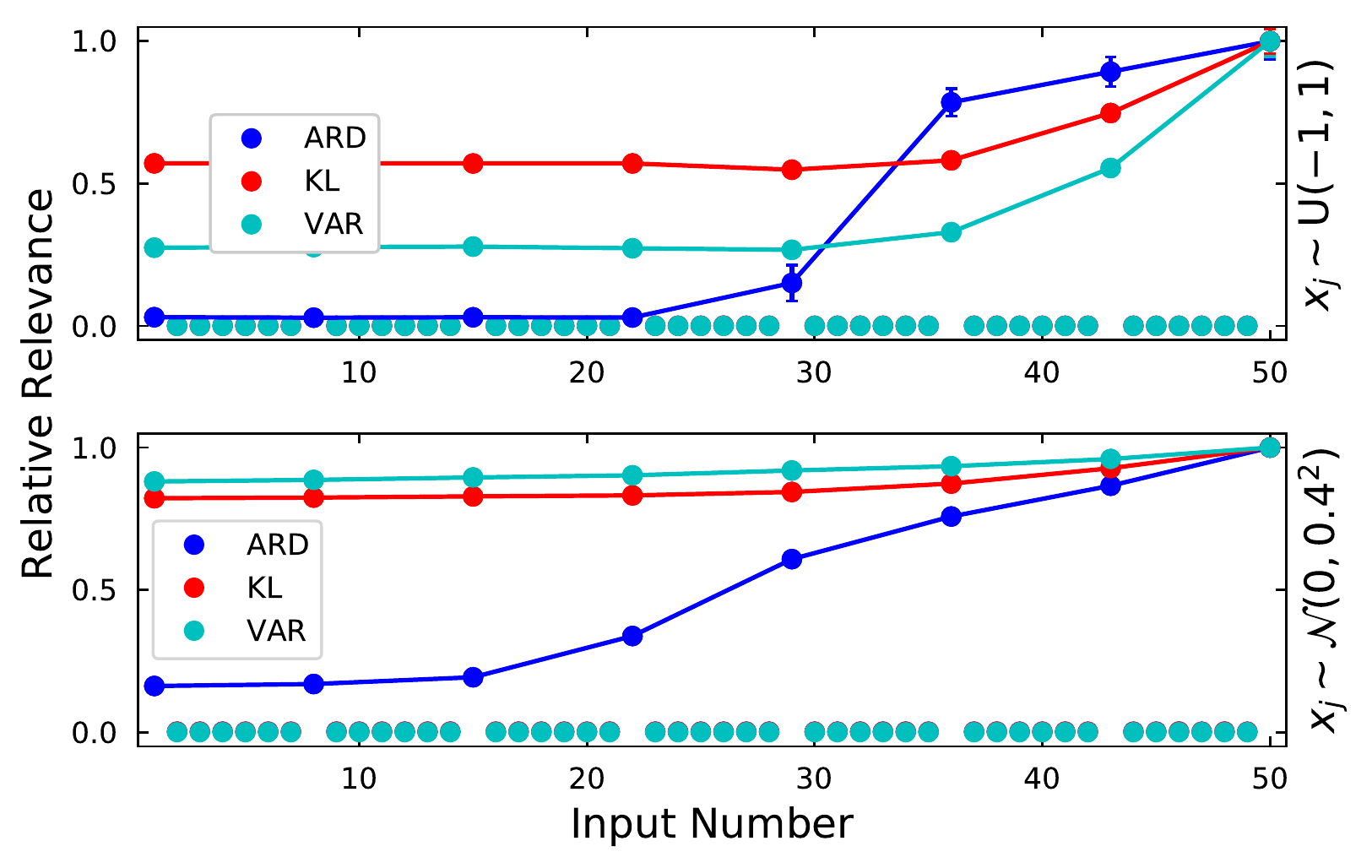}
\caption{Relevance estimates for 50 covariates in the toy model with 8 equally relevant covariates and 42 irrelevant covariates.
The estimates are computed
with ARD (blue), KL (red), VAR (cyan) methods.
The 8 relevant covariates are joined with a line, and range from linear (variable 1) to nonlinear (variable 50).
The results are averaged over 50 data realizations and scaled so that
the most relevant covariate has a relevance of one.} \label{fig:supp1}
\end{figure}

In the toy model presented in the paper, all input variables are equally relevant, thus it does not show how the methods treat irrelevant variables. We also tested an extension of the toy model with 50 variables, 42 of which had no impact on the target variable, and 8
equally relevant with each other. The 8 relevant variables range from linear to nonlinear similarly as in the original toy example in Section~\ref{sec:toymodel}. The relevance values for the 50 variables are
presented in Figure~\ref{fig:supp1}. The results show the same trend as the original toy example, namely that
ARD overly prefers variables with a nonlinear response more than the KL and VAR methods.

\subsection*{Rank One Update of Cholesky Decomposition}

This section presents the method for obtaining the Cholesky decomposition
of a submatrix with one row and one column removed.
This is done by updating the Cholesky decomposition of the full matrix with a rank-one update~\citep{hager1989updating}.
Denote the full matrix and its
Cholesky decomposition as
$\boldsymbol{\Sigma} = \mathbf{L} \mathbf{L}^{\mathsf{T}} \in \mathbb{R}^{p \times p}$.
The goal is to obtain the Cholesky decomposition of the submatrix $\boldsymbol{\Sigma}_{-j,-j}   = \mathbf{L}_{-j,-j} \mathbf{L}_{-j,-j}^{\mathsf{T}}  \in \mathbb{R}^{(p-1) \times (p-1)}$, where the row $j$ and column $j$ are removed
from the full matrix $\boldsymbol{\Sigma}$. A direct Cholesky decomposition
of the submatrix has a computational complexity of $\mathcal{O} (p^3)$, but a
rank one update has only $\mathcal{O} (p^2)$.
If the parts of the lower triangular matrix $\mathbf{L}$ are denoted as
\begin{equation}
\mathbf{L} = \begin{blockarray}{cccc}
        & < j & j & > j  \\
      \begin{block}{c(ccc)}
        < j & \, \mathbf{L}_{A}  & \mathbf{0}    & \mathbf{0}  \, \, \:      \\
        j & \, \mathbf{l}_B^{\mathsf{T}}  & l_{j,j}    & \mathbf{0}^{\mathsf{T}}  \\
        > j & \, \mathbf{L}_{C}  & \mathbf{l}_D    & \mathbf{L}_E     \\
      \end{block}
    \end{blockarray} \in \mathbb{R}^{p \times p}    , 
\end{equation}
The corresponding triangular matrix of the submatrix $\boldsymbol{\Sigma}_{-j,-j}$ is obtained as
\begin{equation}
\begin{split}
\mathbf{L}_{-j,-j} = & \begin{pmatrix}
\mathbf{L}_A    & \mathbf{0}  \, \:        \\
\mathbf{L}_{C}     &  \tilde{\mathbf{L}}_E     \\
\end{pmatrix} \in \mathbb{R}^{(p-1)\times (p-1)} , \\
\tilde{\mathbf{L}}_E \tilde{\mathbf{L}}_E^{\mathsf{T}} = & \, \, \mathbf{L}_E \mathbf{L}_E^{\mathsf{T}} + \mathbf{l}_D \mathbf{l}_D^{\mathsf{T}} .
\end{split} \label{eq:rankone}
\end{equation}
Because $\mathbf{l}_D$ is a vector, the modification to the Cholesky decomposition in
equation~(\ref{eq:rankone}) is a rank-one update.

\subsection*{Additional Predictive Performance Utilities for the Real World Data Sets }

\begin{figure*}[tpb]
  \centering
    \includegraphics[width=0.99\textwidth]{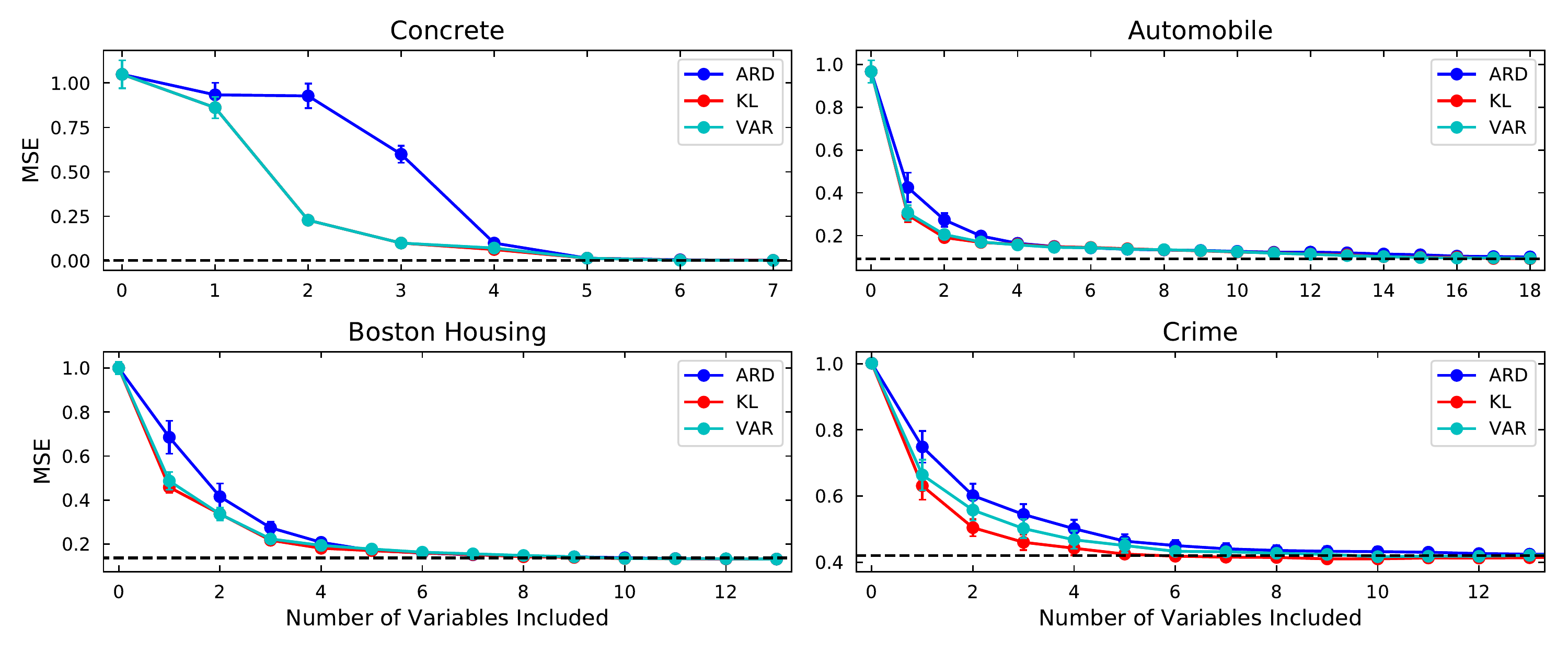}
\caption{Mean squared errors (MSEs) of the test sets with $95 \%$ confidence
intervals for submodels as a function of variables included in the submodel.
Blue depicts variables sorted using ARD, red and cyan depict the KL and VAR methods, respectively. The dashed horizontal line depicts the
MSE of the full model with hyperparameters sampled using the Hamiltonian Monte Carlo algorithm.} \label{fig:allmetrics_regression}
\end{figure*}

\begin{figure*}[tpb]
  \centering
    \includegraphics[width=0.99\textwidth]{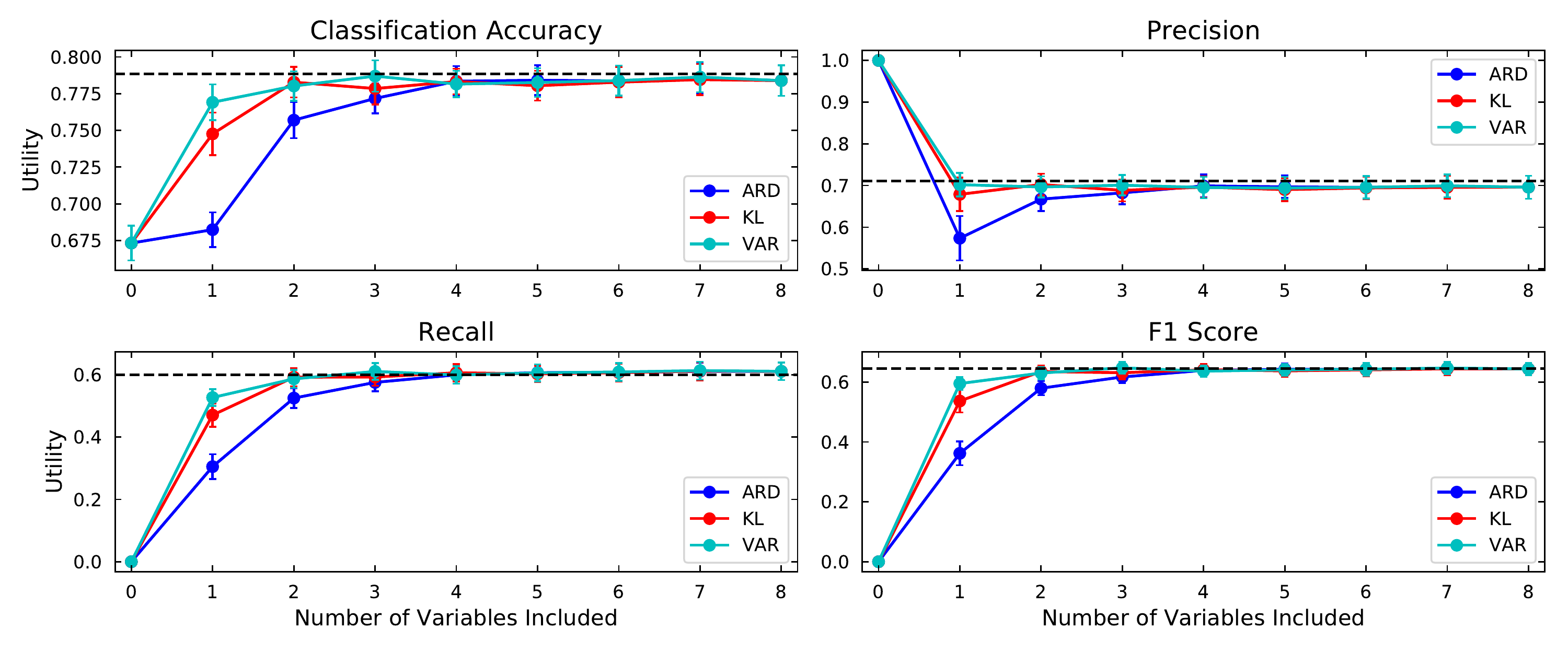}
\caption{classification accuracy, precision, recall, and the F1 score of the test sets of the Pima indians data set with $95 \%$ confidence
intervals for submodels as a function of the number of variables included in the submodel.
The dashed horizontal line depicts the
utilities of the full model with hyperparameters sampled using the Hamiltonian Monte Carlo algorithm.} \label{fig:allmetrics_pima}
\end{figure*}

This section shows the predictive performance of chosen submodels in the real world data sets using different performance utilities. Figure~\ref{fig:allmetrics_regression} is the same as Figure~\ref{fig:allmlpds}, but shows mean squared error instead of mean log predictive density.
Figure~\ref{fig:allmetrics_pima} is the same as Figure~\ref{fig:pima}, but shows classification accuracy, precision, recall, and the F1 score instead of the mean log predictive density.

\end{document}